\documentclass[aps,prx,reprint,10pt,twocolumn,citeautoscript,nofootinbib,superscriptaddress]{revtex4-1}
\pdfoutput=1

\usepackage{natbib}
\usepackage[english]{babel}
\usepackage{letltxmacro}
\usepackage{latexsym}
\LetLtxMacro{\ORIGselectlanguage}{\selectlanguage}
\makeatletter
\DeclareRobustCommand{\selectlanguage}[1]{%
  \@ifundefined{alias@\string#1}
    {\ORIGselectlanguage{#1}}
    {\begingroup\edef\x{\endgroup
       \noexpand\ORIGselectlanguage{\@nameuse{alias@#1}}}\x}%
}
\newcommand{\definelanguagealias}[2]{%
  \@namedef{alias@#1}{#2}%
}
\makeatother
\definelanguagealias{en}{english}
\definelanguagealias{English}{english}
\usepackage{graphicx}
\usepackage{amsmath}
\usepackage{amsfonts}
\usepackage{amssymb}
\usepackage{bm}
\usepackage{color}
\usepackage[percent]{overpic}
\usepackage{soul} 
\usepackage{amssymb}
\usepackage{wasysym}
\usepackage{dsfont}
\usepackage{float}
\usepackage[mathscr]{euscript}
\usepackage{lipsum}
\usepackage{hyperref}
\hypersetup{
    pdfstartview={FitH},    
    colorlinks=true,       
    linkcolor=blue,          
    citecolor=blue,        
    filecolor=magenta,      
    urlcolor=blue           
}

\newcommand{\pdagger}{\phantom{\dagger}}
\newcommand{\be}{\begin{equation}}
\newcommand{\ee}{\end{equation}}
\newcommand{\bea}{\begin{eqnarray}}
\newcommand{\eea}{\end{eqnarray}}

\newcommand{\mc}{\mathcal}

\newcommand{\vect}[1]{\boldsymbol{#1}}

\usepackage{graphicx}
\usepackage[colorinlistoftodos]{todonotes}
\usepackage{verbatim}
\usepackage[normalem]{ulem}

\begin{document}
\title{Itinerant magnetism in Hubbard models with long-range interactions}

\author{Johannes Dieplinger}
\affiliation{Institute of Theoretical Physics, University of Regensburg, D-93040 Germany}
\affiliation{Department of Electrical and Computer Engineering, Princeton University, Princeton, NJ 08544, USA}

\author{Rhine Samajdar}
\affiliation{Department of Physics, Princeton University, Princeton, NJ 08544, USA}
\affiliation{Princeton Center for Theoretical Science, Princeton University, Princeton, NJ 08544, USA}

\author{R. N. Bhatt}
\affiliation{Department of Electrical and Computer Engineering, Princeton University, Princeton, NJ 08544, USA}
\affiliation{Department of Physics, Princeton University, Princeton, NJ 08544, USA}

\begin{abstract}
A wide variety of experimental platforms, ranging from semiconductor quantum-dot arrays to moir\'e materials, have recently emerged as powerful quantum simulators for studying the Hubbard model and its variants.
Motivated by these developments, here, we investigate a generalization of the Hubbard model which includes the effects of long-range Coulomb interactions. 
Working on finite-sized two-dimensional square and triangular lattices, we use exact diagonalization and density-matrix renormalization group calculations to probe the magnetic structure of the ground state in the strong-coupling regime, where $U$ (the onsite repulsion)\,$\gg$ $t$\,(the nearest-neighbor hopping). 
For small electron dopings above the half-filled antiferromagnet, we numerically uncover a rich variety of magnetically ordered states, and in conjunction with theoretical arguments,  infer the phase diagram of the system as a function of doping and interaction strengths. 
In particular, we find that the inclusion of  long-range Coulomb interactions induces an instability of high-spin states---such as the saturated Nagaoka ferromagnet---towards phase separation and stripe ordering. We also present proposals for the observation  of some of our key findings in experiments that would shed further light on this paradigmatic strongly correlated system.
\end{abstract}

\maketitle

\section{Introduction}

Over the last six decades, the Hubbard model \cite{hubbard1964electron,Gut1963,Kan1963} has emerged as a ubiquitous framework to investigate the physics  of systems with strong electronic  correlations. In its original formulation, the model describes interacting spin-$1/2$ electrons hopping on a $d$-dimensional lattice of $N$ sites, and is given by the Hamiltonian,
\begin{equation}
    H^{\pdagger}_0=-t\sum_{[ i, j], \sigma}\left( c^\dagger_{i\sigma}c^{\pdagger}_{j\sigma} +\mathrm{h.c.}\right)+ U\sum_{i}n^{\pdagger}_{i\uparrow}n^{\pdagger}_{i\downarrow}, 
    \label{eq:ham_vanilla}
\end{equation}
where $c_{i\sigma}^\dagger(c^{\pdagger}_{i\sigma})$ are the creation (annihilation) operators for an electron with spin $\sigma$\,$=$\,${\uparrow,\downarrow}$ on a site $i$, $n^{\pdagger}_{i\sigma}\equiv c_{i\sigma}^\dagger c^{\pdagger}_{i\sigma}$ is the electronic density operator, and the sum on $[ i,j ]$  runs over nearest-neighboring sites on a given lattice. While the Hubbard model is exactly solvable by the Bethe ansatz in $d$\,$=$\,$1$ \cite{lieb1968absence}, for higher dimensions, there are only a handful of rigorous analytical results for its ground states \cite{lieb1961two,yamanaka1997nonperturbative,oshikawa2000commensurability,hastings2004lieb,tasaki1998hubbard,li2014exact} owing to the strongly interacting nature of the problem. 
Careful theoretical and numerical studies \cite{Scalapino2007,leblanc2015solutions,schafer2021tracking} have revealed that the Hubbard model hosts a wide variety of complex correlated phases \cite{arovas2022hubbard,qin2022hubbard}, many of which are in close correspondence with observed states in quantum materials.

Motivated primarily by phenomenological similarities to the cuprates \cite{anderson1987resonating,emery1987theory}, much attention has been focused on the possibility of superconductivity in the Hubbard model \cite{dagotto1994correlated, bulut2002dx,qin2020absence,xu2024coexistence}. However, in real materials, superconductivity often appears intertwined with other types of spin and charge order \cite{fradkin2015colloquium}. Hence, it is of fundamental and independent interest to understand the properties of the Hubbard model as a minimal model of itinerant magnetism in dimensions $d>1$. For $d=2$, in particular, on the square lattice, the Hubbard ground state at half filling, $\sum_i\langle n_{i \uparrow} + n_{i \downarrow} \rangle/N =1$ (or equivalently, zero doping), is a N\'eel-ordered insulating antiferromagnet $\forall\, U/t > 0$ \cite{white1989numerical,white2007neel} as a consequence of the perfectly nested Fermi surface. In contrast, on the triangular lattice, the ground state of the undoped Hubbard model for  low $U/t$ is believed to be either a Curie-Weiss metal \cite{merino2006ferromagnetism} or a Luther-Emery liquid \cite{gannot2020hubbard}. On increasing $U/t$ to intermediate values, the system first forms a putative quantum spin liquid state \cite{szasz2020chiral,zhu2022doped,gong2019chiral,hu2019dirac}, before eventually transitioning to a 120$^\circ$-spiral-ordered insulator for larger $U/t$ \cite{huse1988simple, capriotti1999long,szasz2021phase}.

The possibilities for finite doping are even richer. Surprisingly, for the doped Hubbard model, a  rigorous result by \citet{nagaoka1966ferromagnetism} states that for $U$\,$=$\,$\infty$ and non-negative $t$, the ground state on a bipartite lattice endowed with periodic boundary conditions (in $d$\,$\ge$\,$2$) and with a single excess hole away from half filling is ferromagnetic \cite{thouless1965exchange,tasaki1989extension,tian1990simplified}. This is rather remarkable because the undoped system, which is parametrically infinitesimally close to this situation in the thermodynamic limit, provably reduces to an insulating quantum Heisenberg \textit{anti}ferromagnet in the strong-coupling limit \cite{lieb1989two}. 
Although the Nagaoka theorem is exact, its practical relevance is limited, as the prerequisites of infinite onsite interaction strengths and exactly a single dopant are not realistic in experimental setups. Accordingly, much effort has been directed towards understanding whether Nagaoka ferromagnetism survives for 
finite $U/t$ \cite{hanisch1993ferromagnetism,strack1995exact,kollar1996ferromagnetism} and a finite density of carriers  \cite{takahashi1982hubbard, PhysRevB.40.2719, riera1989ferromagnetism, fang1989holes, shastry1990instability, basile1990stability, tian1991stability, putikka1992ferromagnetism, hanisch1995ferromagnetism, wurth1996ferromagnetism, brunner1998quantum, becca2001nagaoka, carleo2011itinerant,ivantsov2017breakdown}.

The formation of ferromagnetic ground states away from half filling can be understood from very general kinetic considerations. Adding dopants creates unoccupied (hole) or doubly occupied (doublon) sites. When these holes or doublons move around on the underlying lattice, they disrupt an antiferromagnetic spin structure, creating ``strings" of misaligned spins that are energetically costly \cite{brinkman1970single,shraiman1988mobile}. In contrast, such hopping does not disturb a ferromagnetic arrangement. As a result, carriers are less confined in a ferromagnetic setting, and the kinetic-energy benefit from their delocalization outweighs the opposing effect of the antiferromagnetic superexchange. This kinetic mechanism leads to the formation of bound states consisting of a dopant surrounded by a cloud of polarized spins; this dressed quasiparticle is referred to as a ferromagnetic polaron \cite{white2001density,maska2012effective}. Although each polaron only carries \textit{local} ferromagnetic correlations, in a system with multiple dopants, the interactions between different polarons can give rise to \textit{global} ferromagnetic order \cite{sb2}, at least on finite lattices.

Such ferromagnetic polarons have been recently observed in experiments on ultracold atoms in optical lattices  \cite{xu2023frustration,lebrat2023observation,prichard2023directly}, which serve as versatile ``quantum simulators''. Another promising platform for quantum simulation of the Hubbard model consists of gate-defined arrays of semiconductor quantum dots \cite{hensgens2017quantum,wang2022experimental}, which can be engineered to generate strong quantum correlations at far lower temperatures (relative to $t$) than achievable with optical lattices. In fact, a few years ago, preliminary signatures of Nagaoka ferromagnetism were observed in a small $2\times 2$ square plaquette of quantum dots \cite{dehollain2020nagaoka}.
However, to accurately describe the physics of such quantum-dot systems, it is important to consider the long-range Coulomb interactions between electrons \cite{wang2011quantum} neglected in the simple Hubbard model \eqref{eq:ham_vanilla}, which focuses on electronic correlations due to local interactions in a
single orbital. Such interactions are also relevant to a host of correlated quantum materials which exhibit ferromagnetism, including twisted bilayer MoTe$_2$ \cite{anderson2023programming}, twisted double-bilayer WSe$_2$ \cite{foutty2023tunable}, MoSe$_2$/WS$_2$ heterostructures \cite{ciorciaro2023kinetic}, and MoTe$_2$/WS$_2$ moir\'e bilayers \cite{tao2024observation}.

These considerations call for the modification of Eq.~\eqref{eq:ham_vanilla} to a long-ranged Hubbard model (introduced in Eq.~\eqref{eq:ham_hub} below), which, to date, has only been studied in certain limited regimes and with---most commonly---up to nearest-neighbor (NN) interactions. In the context of itinerant magnetism though, the effects of the Coulomb interaction on  the ground states of a macroscopic Hubbard system are unclear. In the present study, we address this question and investigate
how the Coulomb repulsion influences the formation and dynamics of magnetic polarons, as well as the interpolaronic interactions which lead to the eventual formation of a fully polarized ferromagnet. Moreover, we demonstrate that instead of uniform magnetic order, the system manifests ordering tendencies towards certain self-organized local inhomogeneities, such as phase separation and stripes. We will reconcile these possibilities in a polaronic picture and also examine if and how these competing orders survive to finite temperatures of relevance to experiments.\\

\section{Model and methods}
\label{sec:model_methods}

To describe systems with long-range Coulomb repulsions, we generalize the standard Hubbard model to
\begin{widetext}
\begin{alignat}{1}
    H =&-t\sum_{[ i, j], \sigma} \left( c^\dagger_{i\sigma}c^{\pdagger}_{j\sigma} +\mathrm{h.c.}\right)+ U\sum_{i}n^{\pdagger}_{i\uparrow}n^{\pdagger}_{i\downarrow} +\frac{1}{2} 
     \hspace*{-0.75cm}
     \sum_{\substack{i\neq j\\ \lvert\lvert \vect{r}_i - \vect{r}_j\rvert\rvert \le r_{\rm cutoff}}} 
     \hspace*{-0.5cm}
     \frac{V}{\lvert \lvert \vect{r}^{}_i- \vect{r}^{}_j\rvert\rvert^\alpha}\,n^{\pdagger}_i n^{\pdagger}_j. 
    \label{eq:ham_hub}
\end{alignat}
\end{widetext}
Here, $\vect{r}_i, \vect{r}_j$ represent the positions of sites $i,j$ on a lattice (in units of the lattice spacing), $n_i \equiv n_{i\uparrow}+ n_{i\downarrow}$ is the total electronic density on site $i$, and $V$ represents the strength of the long-range  interaction, which decays algebraically with an exponent  $\alpha=1$  due to its Coulombic nature. Since $\alpha < d$, the long-ranged nature of the interactions is a relevant perturbation for the long-wavelength physics in the renormalization-group sense \cite{RevModPhys.82.1887}. For computational tractability, we impose a truncation of the Coulomb interaction at a finite cutoff distance, $r_{\rm cutoff}$. This allows us to maintain extensiveness of the interaction when scaling the system size without having to simultaneously scale the strength $V$. 

We are interested in the ground states and finite-temperature properties of this model in two dimensions, and specifically, we study square and triangular lattices with open, cylindrical, or fully periodic boundaries.  Note that the particle-hole symmetry inherent to the Hubbard model \eqref{eq:ham_vanilla} on bipartite lattices is lost due to the long-range interactions, which couple to the charge. Here, we truncate the Coulomb repulsion for distances beyond $\lvert \lvert \vect{r}^{}_i- \vect{r}^{}_j\rvert\rvert > r_{\rm cutoff}=2$, which translates to retaining density-density interactions between sites that are up to third-nearest neighbors (3NNs) apart for the geometries considered. Throughout the rest of this work, we will refer to Eq.~\eqref{eq:ham_hub} with this particular choice of truncated interactions as the ``long-ranged Hubbard model''. This approximation preserves the dominant short- to intermediate-range contributions of the $1/r$ interaction that govern the electrostatic redistribution of charge, while keeping the Hamiltonian numerically manageable within matrix-product-operator (MPO) representations. In particular, retaining interactions up to this range captures the leading energetic competition between kinetic delocalization and Coulomb repulsion that underlies the phase separation and stripe phenomena discussed below. Extending the interaction to longer distances would quantitatively increase the electrostatic penalty for macroscopic charge segregation, but is not expected to alter the qualitative mechanism driving any instability of ferromagnetism in this regime.

To begin, let us consider the case when the interactions in the second line of Eq.~\eqref{eq:ham_hub} are restricted to only nearest neighbors ($\lvert \lvert \vect{r}^{}_i- \vect{r}^{}_j\rvert\rvert =1$); this approximated Hamiltonian is often called the ``extended Hubbard model''. At half filling, the phase diagram of this model has been extensively studied  in $d=1$ as a function of  $\vartheta = V/U$. A variety of weak- and strong-coupling studies show that  the system is in a charge-density wave (CDW) phase if $\vartheta$ is large and in a spin-density wave (SDW) phase for small $\vartheta$, with the transition between them occurring around $\vartheta \simeq 1/2$ \cite{voit1992phase, van1994extended, lin1995phase, zhi2002critical,tanaka2005effects, benthien2007spin}. 
In between these two density-wave orders, there also exists an intermediate bond-ordered phase \cite{tsuchiizu2004ground,ejima2007phase,glocke2007half}. In two dimensions, at or near half filling on the square lattice, the extended Hubbard model is known to host symmetry-broken CDW and SDW phases \cite{yamamoto1992broken,yan1993theory, aichhorn2004charge, sherman2023two}, in addition to states with orbital
antiferromagnetism or spin-nematic order \cite{chattopadhyay1997phase}, and $s$- or $d$-wave superconductivity \cite{zhang1989extended, laad1991extended, huang2013unconventional, chen2023superconducting}.

In this work, we examine the magnetic ground states of the \textit{long-ranged Hubbard model} for different  parameters $U/t,V/t$ and electron dopings $\delta \equiv \sum_i \langle n_i \rangle/N -1 > 0$ away from half filling in the strongly correlated regime of $U/t \gg 1$. To do so, we adopt a two-pronged approach. Firstly, we study small $4\times 4$ plaquettes, which can be diagonalized exactly by means of Lanczos methods making use of translational and reflection symmetries along with particle-number and $S^z$ conservation. Next, we set up much larger samples with cylindrical boundary conditions, i.e., periodic in the $\hat{y}$-direction and open along the $\hat{x}$-axis. These systems can be efficiently studied using methods based on matrix product state (MPS) ans\"atze. Specifically, to obtain the ground state and its correlators, we use the density-matrix renormalization group (DMRG) algorithm \cite{stoudenmire2012studying}. The numerical errors of this method are controlled by the maximum bond dimension $\chi$ and the truncation error $\varepsilon$ of the associated singular value decomposition. Iterating over sweeps with progressively increasing bond dimensions, up to $\chi  =2400$, we use a threshold of $\varepsilon <10^{-7}$ as our criterion for convergence. 

The exact diagonalization results presented in this work are necessarily restricted to small clusters and serve primarily as benchmarks and illustrative examples. Our central conclusions regarding the stability of ferromagnetism, phase separation, and stripe formation are based on simulations of the larger systems using tensor networks on cylindrical geometries. These geometries provide a controlled compromise between capturing two-dimensional physics and maintaining numerical tractability within MPS methods. We note that small clusters with open boundary conditions can exhibit strong finite-size and boundary effects; accordingly, we avoid drawing conclusions about the thermodynamic limit based solely on such systems.

In either approach, for each computed wavefunction, we measure all spin-spin correlation functions, which give us access to the local spin structure and reveal potential magnetic orderings. Such a two-point function is defined by
\begin{equation}
    \mathcal{C}_{ij}=\langle \mathbf{S}_i\cdot \mathbf{S}_j \rangle, 
    \label{eq:C}
\end{equation}
where $\mathbf{S}^{\pdagger}_i \equiv c^\dagger_{i\mu} \vect{\tau}^{\pdagger}_{\mu \nu} c^{\pdagger}_{i \nu}$ for Pauli matrices $\{ \tau^{\upsilon} \}, \upsilon = 1,2, 3$. 
The total spin of the system can  be expressed as the sum over all matrix elements of the spin-spin correlator as
\begin{equation}
    {\textbf{S}}_\mathrm{tot}^2=\sum_{i,j=1}^N\mathcal{C}_{ij} \equiv S(S+1),
\end{equation}
where $S$ is the net spin quantum number.

Besides the zero-temperature ground states, we will also be interested in exploring finite-temperature properties, which are relevant to all experimental systems. In fact, for many cold-atom quantum simulators, the effective temperatures of the prepared Fermi-Hubbard models are only somewhat smaller than the kinetic energy of a single electron and may easily exceed the spin-exchange coupling $J \sim t^2/U$. It is therefore important to ask if and to what extent the magnetic structures of the ground states, as obtained with DMRG, survive thermal fluctuations at large effective temperatures. However, this is a challenging task because finite-temperature calculations require access to the full spectrum---not just the low-lying eigenvalues---which is out of reach with exact diagonalization for even modest system sizes of $20$--$40$ electrons. 

In our study of the  long-ranged Hubbard model, we address this issue using minimally entangled typical thermal states (METTS) \cite{stoudenmire2010minimally}, which are a recent addition to the family of MPS-based methods. 
A so-called typical thermal state can be defined as the imaginary-time evolution, up to an imaginary time  $\beta$\,$=$\,$1/T$, of a classical product state (CPS). 
The expectation value of any observable of interest is then computed  using this thus-evolved pure state, whereafter the next CPS is obtained by collapsing the state by a projective measurement. 
Iterating this procedure, we can systematically prepare an arbitrary number of such METTS, and the finite-temperature trace of an observable is evaluated by averaging over these random realizations. In principle, the wavefunction collapse at the end of each iteration cycle to obtain the new CPS can be carried out in any basis. This choice  of basis does not affect the physical outcome but does influence the correlations between successive METTS, which need to be minimized for optimal efficiency. Here, we choose the off-diagonal $X$ basis of the spin vector for computing the finite-temperature correlation functions.

All the DMRG and METTS calculations reported here are explicitly U(1)-symmetric and conserve particle number (i.e., the total $S^z$). Our ground-state calculations are typically performed in the $S^z=0$ sector near half filling; however, to probe and verify ferromagnetic states, we also worked in sectors with larger $S^z$ (including $S^z = S_{\rm max}$) and initialized calculations with polarized product states where appropriate. To characterize the total spin of the system, in principle, one can compute $\mathbf{S}_{\rm tot}^2 = \sum_{ij}\langle \mathbf{S}_i \cdot \mathbf{S}_j\rangle$ directly from the converged matrix-product state. While such global quantities can be evaluated accurately within DMRG, in practice, reliable convergence---particularly for quasi-two-dimensional geometries---requires careful control of bond dimension and truncation error. We therefore performed convergence checks with increasing bond dimension, benchmarked against exact diagonalization on small clusters where available, and mostly rely on the long-range spin correlations (which can be evaluated more robustly) as a diagnostic tool.

\section{Magnetic properties on the square lattice}

On the square lattice, calculations on the extended Hubbard model---with only NN repulsion---have shown that ferromagnetism can arise as a result of the coexistence of CDW and SDW orders \cite{murakami2000possible}. For true Coulomb interactions (beyond just NNs) in one dimension, it has been suggested that the long-ranged character of the potential is important to stabilize ferromagnetism \cite{farkavsovsky2014ferromagnetism}. However, in $d$\,$=$\,$2$, the corresponding ground-state properties of the long-ranged Hubbard model \eqref{eq:ham_hub} for finite carrier dopings, and the relationship among various magnetic orders, have yet to be understood. This will be our focus in the following discussion.

\subsection{Square clusters}

\begin{figure}[t]
    \centering
    \includegraphics[width=\linewidth]{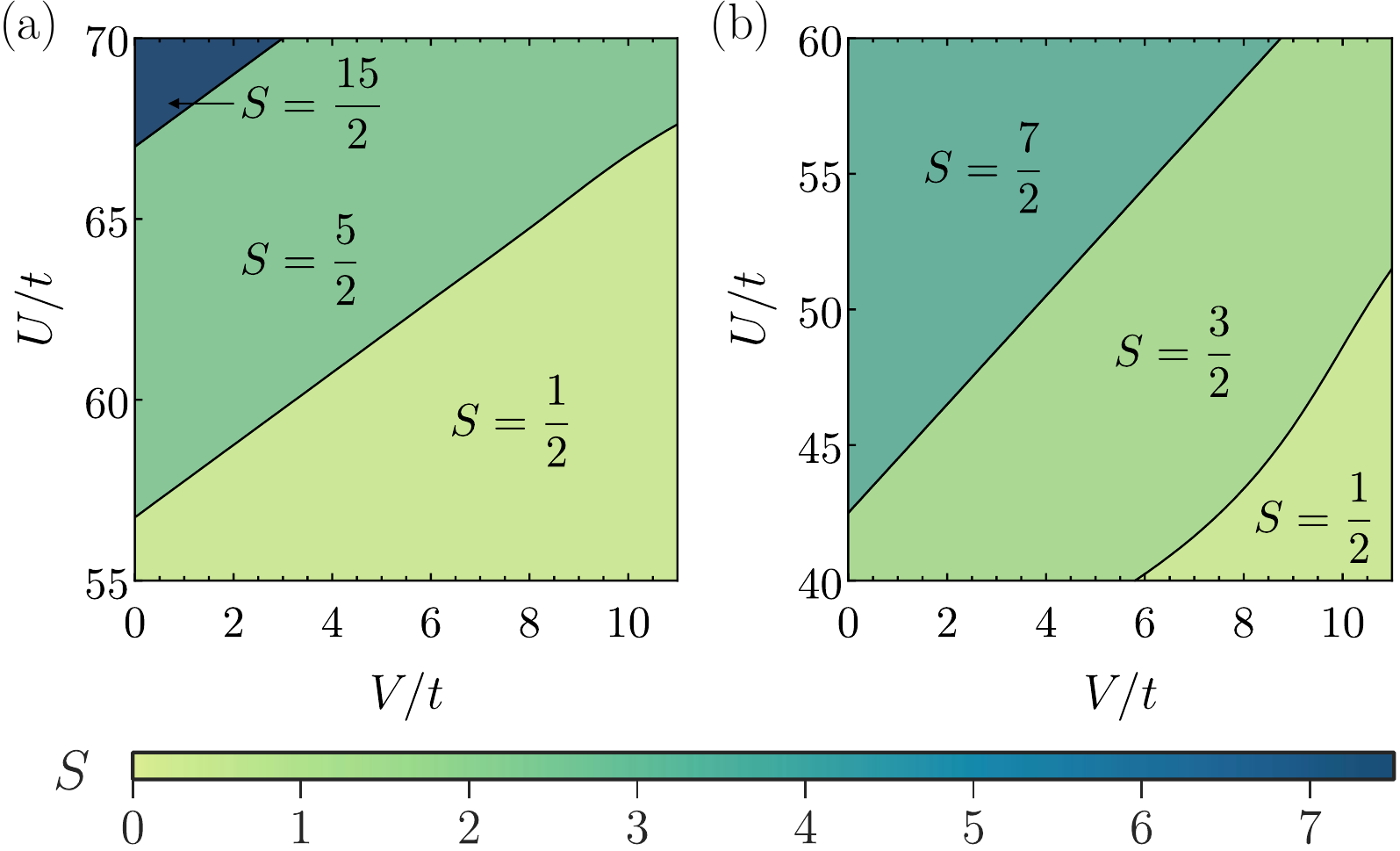}
    \caption{Total spin $S$ of a $4\times4$ square lattice with periodic boundary conditions, as a function of the  interaction strength $V/t$ and the onsite repulsion $U/t$. The samples are doped with (a) a single electron, and (b) three excess electrons  above half filling. Data obtained by exact diagonalization (ED) using the Lanczos algorithm.}
    \label{fig:sq_plaq_ED}
\end{figure}

Motivated by recent experiments on interacting quantum dot arrays \cite{dehollain2020nagaoka,wang2022experimental}, we begin by considering a $4\times4$ cluster of the long-ranged Hubbard model on the square lattice. Although a $16$-site system is relatively small, as we will see, it proves useful in identifying certain key features that we can then generalize to larger system sizes in subsequent sections. We compute the spin-spin correlation functions \eqref{eq:C} of this system using exact diagonalization. Figure~\ref{fig:sq_plaq_ED} shows the total spin $S$ of the $4\times 4$ arrays with periodic boundaries as a function of the onsite and Coulomb interactions $U/t, V/t$, for different dopings close to half filling.

For sufficiently large $U/t$, we expect the ground state to be a saturated Nagaoka ferromagnet. Indeed, for a single dopant electron above half filling [Fig.~\ref{fig:sq_plaq_ED}(a)], we observe a change in the ground state from an antiferromagnetic spin-zero configuration to a fully polarized high-spin (for the present system, $S=15/2$) state as $U/t$ is increased in the standard Hubbard model ($V=0$). Between these two extremal cases, there exists an intermediate partially polarized phase in which the ground state carries spin $S=5/2$. 

Switching on the long-range Coulomb interactions next, we find that the qualitative picture remains the same;  for a given $V> 0$, the system continues to exhibit two transitions as a function of increasing $U/t$: first, from a minimal-spin to a partially polarized state, and then, to a fully polarized ferromagnet. Interestingly, the critical $U/t$ required for the formation of the saturated ferromagnet shifts to higher values when increasing $V/t$, indicating that the repulsive interaction hinders ferromagnetism of the Nagaoka type and instead favors lower-spin states. This can be understood in a polaronic picture. Consider a polaron comprised of a circular bubble of polarized spins around the doublon with the spins further away antiferromagnetically aligned. In the limit of $U/t \rightarrow \infty$, where the Hubbard model \eqref{eq:ham_vanilla} reduces to the so-called $t$-$J$ model, the radius of the polaron in the absence of any long-range interactions scales as $R \approx 1.12 J^{-1/4}$ \cite{white2001density}, where $J$ is the strength of the antiferromagnetic superexchange. The introduction of the $1/r$-decaying interactions on top leads to a correction to the polaron's extent, which, to leading order, is given by $\delta R \propto - V/J + \mc{O}(V^2)$. Now, the transition to the Nagaoka ferromagnet occurs when the area of the polaron ($\sim \pi R^2$) grows to be of the order of the system size, but due to the negative sign of the Coulombic contribution $\delta R$, for a fixed $J$, the radius of the polaron is \textit{smaller} than it would be in the absence of the intersite repulsion. Consequently, a larger $U/t$ is required to fully polarize the system, as is indeed the case in Fig.~\ref{fig:sq_plaq_ED}(a).

On increasing the doping to two excess electrons (not shown), surprisingly, we find that there is no transition to a high-spin state with increasing $U/t$, at least over the parameter range studied (up to $U/t \le 1000$). A possible explanation of this observation is as follows. The two doublons form two ferromagnetic polarons, both of which have a finite spin and, for low $U/t$, a finite extent which is not large enough to polarize the full sample. Then, as $U/t$ is increased (or equivalently, $J$ is reduced), the polarons grow in size and their wavefunctions overlap in real space. However, the polaronic clouds associated with the two doublons do not align but instead form a singlet, resulting in a net spin-zero state as observed. This hypothesis is also consistent with the behavior of the system when doped with three excess  electrons above half filling [Fig.~\ref{fig:sq_plaq_ED}(b)]. According to this consideration, the three polarons  now formed try to antialign, i.e., two of them form a singlet while the third one is free and contributes to the total spin. Therefore, in Fig.~\ref{fig:sq_plaq_ED}(b), which plots the phase diagram for doping with three electrons, we observe a similar picture as for a single dopant: a transition from a spin-zero state to a high-spin state via an intermediate low-spin state. However, the high-spin state is not fully polarized---it has a total spin of only $S=7/2$---supporting the hypothesis of polarons forming singlets pairwise.

\begin{figure}[b]
    \centering
    \includegraphics[width=\linewidth]{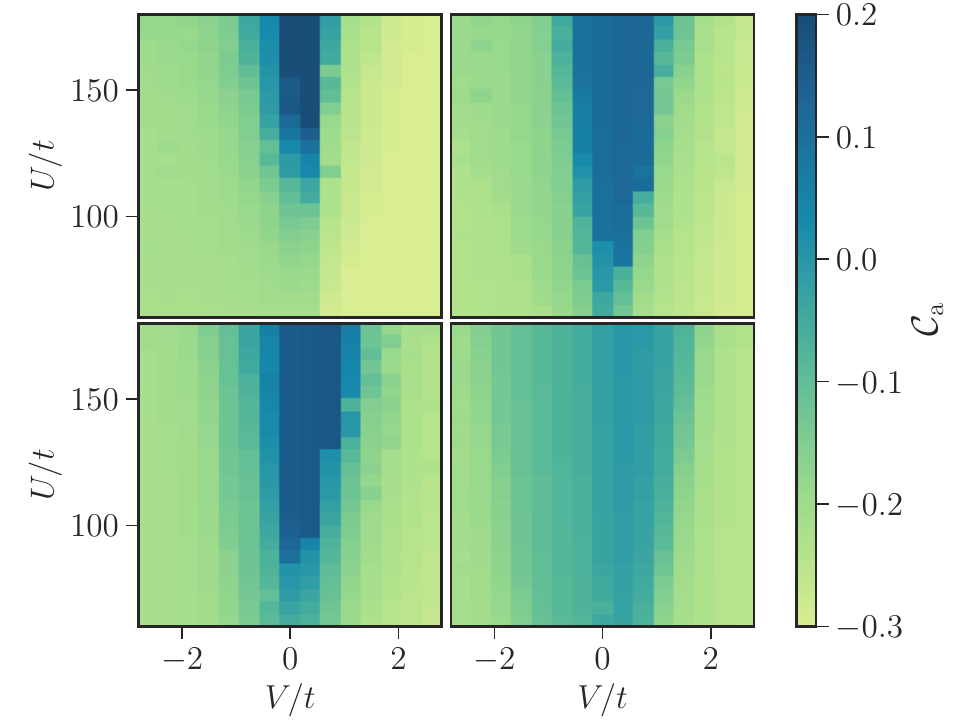}
    \caption{Nearest-neighbor correlator $\mathcal{C}_\mathrm{a}$ as calculated for the ground state of the long-ranged Hubbard model~\eqref{eq:ham_hub}, as a function of $U/t$ and $V/t$. By definition, $\mathcal{C}_\mathrm{a}$ is negative for antiferromagnetic correlations and positive for ferromagnetic ones. The panels correspond to dopings of one (upper left), two (upper right), three (lower left), and four (lower right) excess electrons above half filling in a $4\times 4$ sample with open boundary conditions. Data obtained by DMRG with maximum bond dimension $\chi=2400$.}
    \label{fig:sq_open_DMRG}
\end{figure}

Next, we consider the same $4\times4$ cluster, but with open boundary conditions.
This breaks translational symmetry, making exact diagonalization unfeasible for multiple dopants; we thus turn to DMRG simulations. As discussed in Sec.~\ref{sec:model_methods}, the total spin $\textbf{S}^2_\mathrm{tot}$ can in principle be evaluated from the converged matrix-product state, but its accurate determination on quasi-2D geometries can require substantially larger bond dimensions than purely local observables. We therefore focus here on the local nearest-neighbor spin correlation $\mathcal{C}_\mathrm{a}$\,$\equiv$\,${1}/{\mc{N}_{b}}\cdot \sum_{[ i,j] }\mathcal{C}_{ij}$, where $\mc{N}_{b}$ is the number of NN bonds on the lattice. We use this quantity, which is negative for an antiferromagnet and saturates to its maximal possible value in the fully polarized ferromagnetic phase, to  diagnose the magnetic structure of the ground state. We emphasize that $\mathcal{C}_\mathrm{a}$ serves primarily as an efficient local indicator for large-scale parameter sweeps, as it converges rapidly with bond dimension. To confirm the nature of the magnetic state, we complemented this diagnostic with calculations of the total spin $\mathbf{S}^2_{\rm tot}$ and longer-range spin--spin correlators. These additional checks ensure that the identification of ferromagnetic or non-ferromagnetic phases does not rely solely on nearest-neighbor correlations.

In Fig.~\ref{fig:sq_open_DMRG}, the averaged NN correlation $\mathcal{C}_\mathrm{a}$ is shown as a function of $U/t$ and $V/t$. Qualitatively, the overall picture is similar to that in Fig.~\ref{fig:sq_plaq_ED}. Finite repulsive interactions hinder the formation of the Nagaoka state; if the state is ferromagnetic at $V$\,$=$\,$0$ to begin with, it is eventually destroyed for large enough $V$. 
However, there are some important quantitative differences compared to the case with periodic boundaries. First, the threshold value of $V/t$ beyond which ferromagnetic correlations are suppressed is significantly lower than in the case with periodic boundaries. We attribute this to the fact that due to the reduced coordination number at the boundaries, excess charge tends to accumulate at the edges of the sample. This results in a corresponding depletion of charge in the bulk, rendering its charge density to be at nearly half filling, which favors an antiferromagnetic spin texture.

Although the repulsive Coulomb potential corresponds to a positive $V/t$, for completeness, it is also interesting to consider the case of attractive interactions, $V$\,$<$\,$0$. Such a situation is realizable in ultracold fermionic gases \cite{esslinger2010fermi} and certain narrow-band materials \cite{micnas1990superconductivity}, and has been much explored in the context of superconductivity  in the extended Hubbard model \cite{micnas1988extended,su2004phase,peng2023enhanced, sousa2024half}. 
In Fig.~\ref{fig:sq_open_DMRG}, we observe that the phase diagram is similar for both negative and positive $V/t$ in the sense that attractive interactions also shift the critical $U/t$ for ferromagnetism to larger values (cf.~Ref.~\onlinecite{lhoutellier2015fermi}).  This phenomenological observation for $V$\,$<$\,$0$ has its roots in phase separation: attractive interactions naturally favor the clustering of charge and lead to the formation of electron-rich and electron-deficient (or hole-rich) regions, thereby destabilizing ferromagnetism. Similar behavior has also been identified in the extended Hubbard model both at and away from half filling \cite{su2004phase,fresard2016charge,van2018extended}.

\subsection{Phase separation and stripe orders}

Our calculations for the $4\times 4$ square cluster indicate an instability of the ferromagnetic state towards low-spin configurations on account of the long-range interactions. The consistency of these results with the polaronic framework also suggests their direct generalization to extended two-dimensional systems, which we now investigate.

We next study the long-ranged Hubbard model on the square lattice in rectangular (quasi-2D) geometries with widths $W$ and lengths $L \gg W $, which can be efficiently simulated with DMRG. Unless specified otherwise, we work with cylindrical boundaries, which are periodic along the width of the cylinder and open along the length.

\begin{figure}[t]
    \centering
    \includegraphics[width=\linewidth]{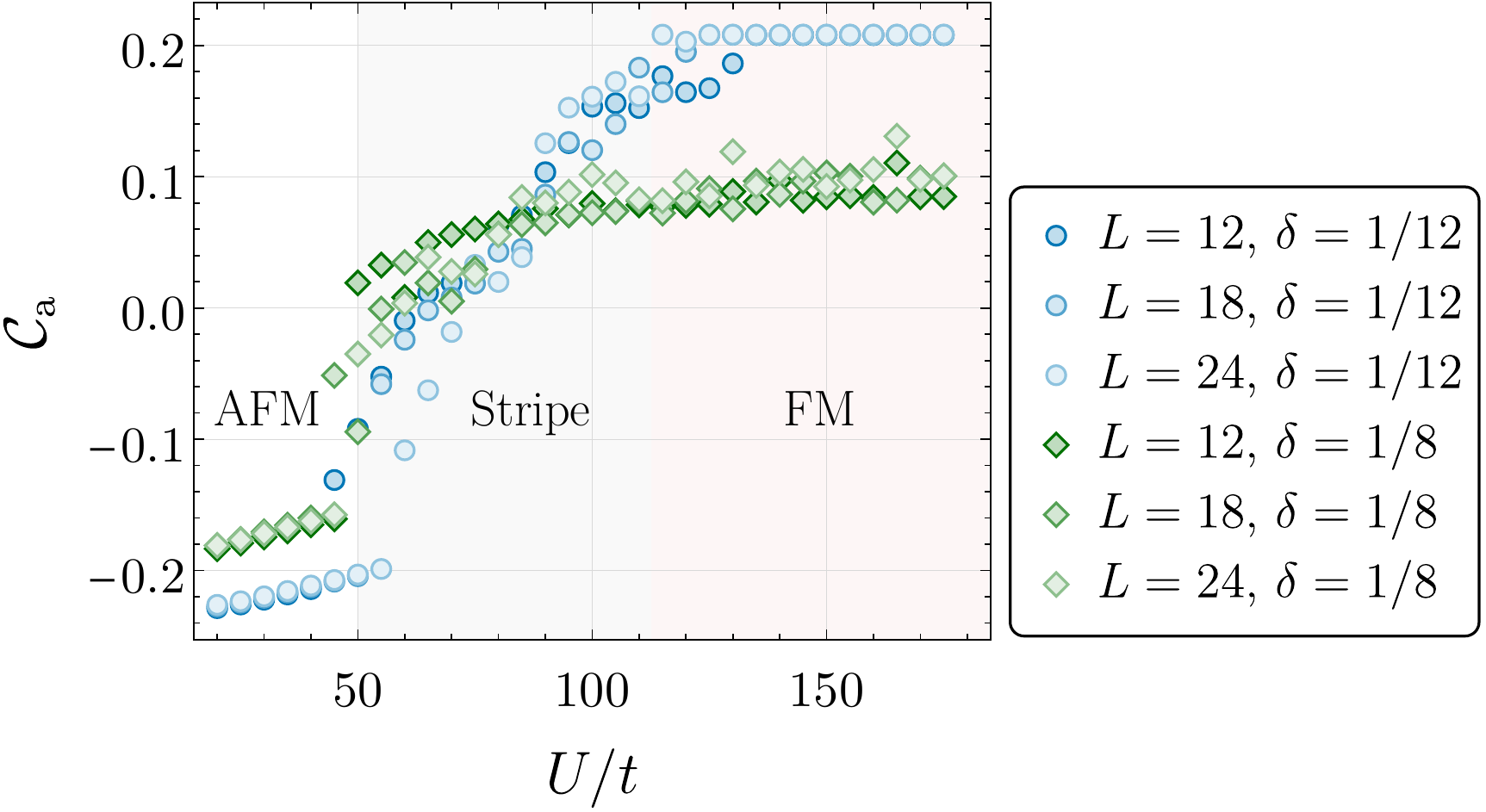}
    \caption{Nearest-neighbor correlator $\mathcal{C}_\mathrm{a}$ (negative for antiferromagnetic correlations, positive for ferromagnetic) as a function of the onsite repulsion $U/t$ on a width-4 cylinder of length $L=12$, $18$, $24$ with doping fractions $\delta=1/12$ and $1/8$ above half filling, shown here for the pure Hubbard model---without long-range interactions---on a square lattice. Data obtained by DMRG with maximum bond dimension $\chi=2400$.}
    \label{fig:sq_cylinder_Uvar}
\end{figure}


Before introducing the long-range Coulomb interactions, we have to ascertain the existence and range of high-spin ground states of the ordinary Hubbard model \eqref{eq:ham_vanilla} on such cylinders. Figure~\ref{fig:sq_cylinder_Uvar} shows the average NN spin correlator for $V$\,$=$\,$0$, as a function of the onsite repulsion $U/t$ for various system sizes with doping fractions $\delta$\,$=$\,$1/12$ and $1/8$ above half filling; these electron concentrations are chosen because they were empirically found to be close to optimal for ferromagnetism \cite{sb2}. 
Once again, we observe an antiferromagnetic ground state up to around $U/t \lesssim 50$, after which the onset of Nagaoka ferromagnetism is apparent from the saturation of $\mathcal{C}_\mathrm{a}$. Note that the critical onsite interaction strength required to produce a nonzero-spin state is considerably lower than for the $4\times 4$ clusters examined earlier. This reduction stems from the increased mobility of the ferromagnetic polarons on the extended cylinder compared to the smaller cluster, where polarons are always localized near the center of the sample \cite{white2001density,sb2} (recall that the mechanism driving ferromagnetism is the kinetic energy gained from delocalization).

\begin{figure}[t]
    \centering
    \includegraphics[width=\linewidth]{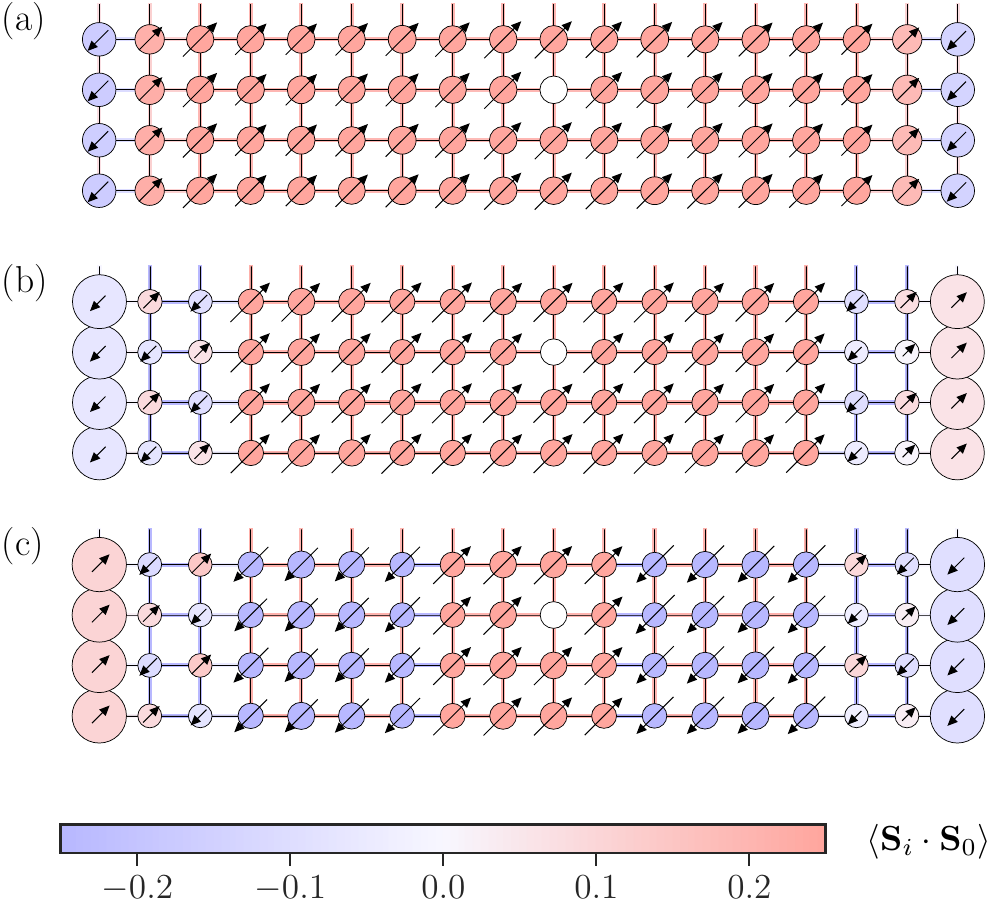}
    \caption{Visualization of spin-spin correlations and onsite electronic densities for the long-ranged Hubbard model in rectangular samples with varying $V/t=$ (a) 0.4, (b) 3.6, and (c) 6.0  at $U/t=175$. The size of the lattice points is proportional to the local excess electron density $\langle n_i \rangle -1$ in the ground state. The color
of the circles as well as the orientation of the arrows indicates the sign of the spin correlations $\langle \mathbf{S}_i\cdot \textbf{S}_0\rangle$ between the spin at any given site $\textbf{S}_i$ and a central reference site (marked in white). In this convention,  red (blue) represents positive (negative) correlations with the magnitude thereof conveyed by the length of the arrows. Data obtained by DMRG with maximum bond dimension $\chi=2400$.}
    \label{fig:sample_square}
\end{figure}

Next, we add in the long-range repulsions to this landscape. In Fig.~\ref{fig:sample_square}, the evolution of the ground state with increasing $V$ is shown for an $18\times 4$ cylinder with $\delta = 1/8$, starting from the ferromagnetic state at large $U/t$. For a small finite $V/t$\,$=$\,$0.4$, we find that the bulk of the sample forms a single large ferromagnetic domain; it is only at the edges  that the correlations flip sign. This is because, as mentioned above, the fully polarized ferromagnet becomes energetically less favorable for $V > 0$, and the system partially compensates for this with the superexchange energy gained across the two domain walls between the bulk and the 1D ferromagnets at the boundary. 

Increasing the interaction strength to larger values, $V/t=3.6$, interestingly, we observe a clear phase separation of the system  into ferromagnetic and antiferromagnetic regions. While the charge accumulation at the boundaries is prominent, immediately adjacent to the edges, we see the development of a low-density region which exhibits antiferromagnetic correlations. In contrast, the bulk remains ferromagnetic with only a slight modulation of the charge along the longitudinal direction. This phase separation can be understood from a classical electrostatic description of the repulsive interactions, which lead to a redistribution of the charge density. We have shown in earlier work that in the regular  Hubbard model ($V$\,$=$\,$0$), the critical value of $U/t$ for the transition to the Nagaoka state varies nonmonotonically with the doping fraction \cite{sb2}. In particular, let  $\delta^*$ be the optimal doping  that minimizes the $U/t$ needed to realize a ferromagnetic phase. Then, on going to dopings slightly above or below this optimal value, for a fixed $U/t$, the system may be driven into an antiferromagnetic phase \textit{even if} it were originally fully ferromagnetic for $\delta = \delta^*$. Such a scenario is induced by the collective (but classical) reorganization of the charge distribution by the repulsive interaction. In the vicinity of the edges, due to the charge depletion,  the effective local doping fraction falls below that required for ferromagnetism, thereby converting the region to a locally antiferromagnetic configuration.

It is worth noting that part of the shift in the critical interaction strength with increasing $V$ can already be understood within the extended Hubbard model with nearest-neighbor interactions, where $V$ renormalizes local energetics and modifies the effective exchange scale. However, interactions beyond nearest neighbors play an essential role in determining the energetics of spatially separated charge configurations. In particular, longer-range terms significantly influence the interaction between ferromagnetic polarons and the electrostatic cost of charge inhomogeneity, thereby affecting whether the system favors macroscopic phase separation or modulated states such as stripes. While nearest-neighbor interactions capture certain local effects, the inclusion of longer-range tails is crucial for describing the emergence and structure of these spatially extended phases.

\begin{figure}[t]
    \centering
    \includegraphics[width=0.9\linewidth]{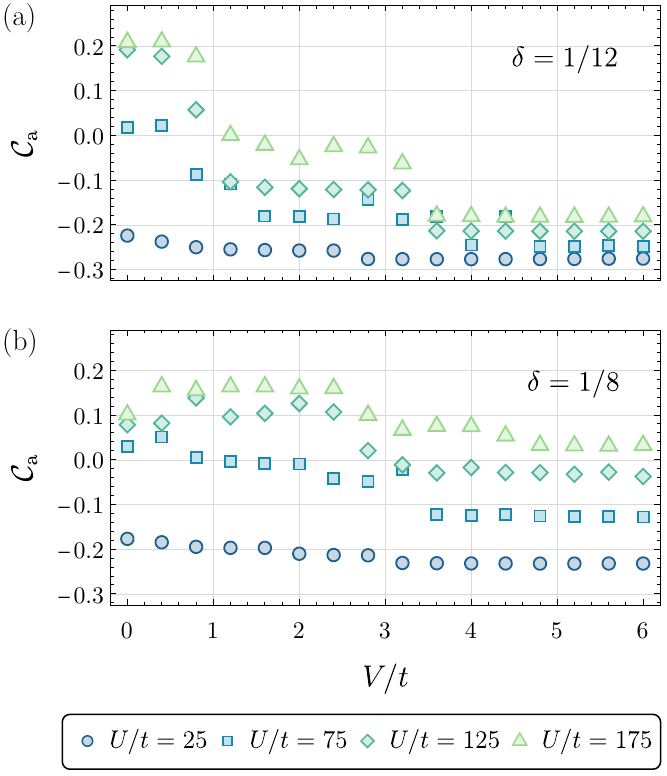}
    \caption{Nearest-neighbor spin correlation $\mathcal{C}_\mathrm{a}$ in the square-lattice long-ranged Hubbard model as a function of the repulsive interaction strength $V$ for several combinations of $U/t$ and $\delta$. The samples have the same geometry as in Fig.~\ref{fig:sample_square}. Data obtained by DMRG with maximum bond dimension $\chi=2400$.}
    \label{fig:square_V_phase}
\end{figure}

\begin{figure}[t]
    \centering
        \includegraphics[width=0.9\linewidth]{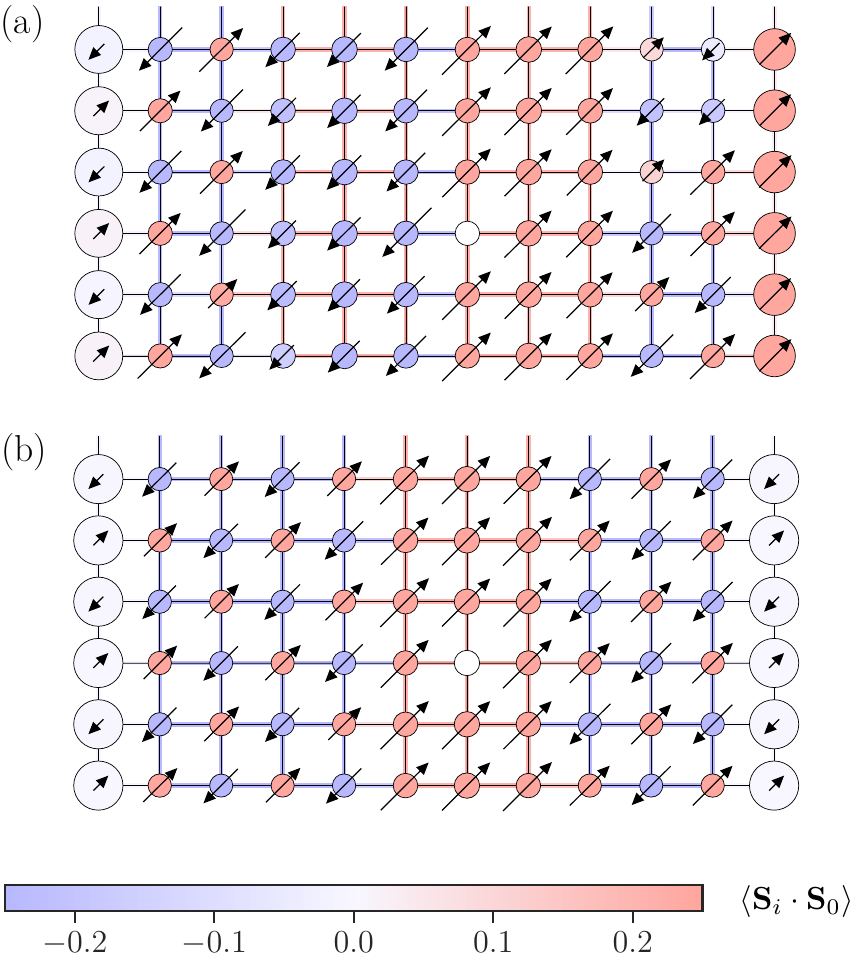}
     \caption{Visualization of spin-spin correlations and electronic densities as in Fig.~\ref{fig:sample_square}, but for a different (wider) geometry.  The configurations here are  plotted for $V/t=$ (a) 2.8, and (b) 5.6 at $U/t=175$. Data obtained by DMRG with maximum bond dimension $\chi=2400$.}
    \label{fig:sample_square_Ny6}
\end{figure}

Finally, for even stronger Coulomb repulsions, $V/t=6.0$, the spin texture of the central region is also fundamentally modified and we observe the development of stripe order, i.e., a uniaxial modulation of the spin density. We can understand the origin of the stripes by considering the competition between three factors: the antiferromagnetic exchange, which favors the spontaneous generation of domain walls, the Coulomb repulsion, which tends to localize charge, and the kinetic  term in the Hamiltonian which prefers the delocalization of electrons \cite{emery1999stripe}.  The microscopic compromise between these three energy scales determines whether a fully polarized ferromagnetic or a striped pattern dominates in the bulk.

Filling in between the discrete values of $V/t$ examined in Fig.~\ref{fig:sample_square},  we plot the effect of $V$ on the nearest-neighbor average spin correlations   in Fig.~\ref{fig:square_V_phase}. The trends discussed above are also reflected in $ \mathcal{C}_\mathrm{a}$, and in general, increasing $V/t$ reduces the magnitude of the ferromagnetic correlations. These observations are also stable with respect to  the length of the cylinder, as verified for $L=12$, $18$, and $24$, so our numerics are not limited by finite-size effects in this direction.

Scaling up the size of the system in the transverse direction is a much more challenging task. Nonetheless, we repeat our calculations above for a width-6 cylinder, which is close to the limit of today's state-of-the-art DMRG methods. In Fig.~\ref{fig:sample_square_Ny6}, we show that our earlier observations of phase separation for low $V/t$ followed by the onset of stripe ordering at larger $V/t$ persist on these wider cylinders as well. Although we cannot systematically perform finite-size scaling analyses given the limitations on the cylinder width, it is not unreasonable to expect that these general features outlined here will continue to hold for large samples.


\section{Kinetic magnetism on triangular lattices}

On the square lattice, the Nagaoka state arises at large $U/t$ solely from the balance between the weakening superexchange $J$\,$\sim$\,$t^2/U$ and the kinetic energy of the dopant $\sim$\,$\delta\, t$. When considering the influence of the Coulomb repulsion on square-lattice Hubbard ferromagnetism, we are therefore always in a regime where $V \gg J$. Interestingly, however, on the triangular lattice, we can have an interaction-induced transition out of the ferromagnetic phase without such a separation of scales, i.e., when $V \sim J$, as we show next.

In triangular lattices, itinerant magnetism is closely connected to the notion of kinetic frustration \cite{morera2022hightemperature,schlomer2023kinetictomagnetic,samajdar2023nagaoka}. Owing to the nonbipartite nature of the lattice, the motion of a single hole (doublon) within a spin-polarized environment results in destructive (constructive) quantum interference between various paths \cite{haerter2005kinetic}. This has important energetic consequences: the minimum kinetic energy of a single hole on a triangular lattice is $-3\, t$ in a uniform ferromagnetic background, whereas that of a doublon is twice that,~$-6\, t$  \cite{zhang2018pairing}. To reduce their kinetic energy as much as possible, moving holes therefore tend to foster antiferromagnetic spin correlations around themselves, thereby alleviating the frustration \cite{sposetti2014classical,lisandrini2017evolution,zhang2018pairing}. Conversely, doublons promote a local ferromagnetic environment \cite{hanisch1995ferromagnetism,kanasz2017quantum}. As a result of this kinetic frustration, the critical onsite interaction strength for ferromagnetism is much reduced relative to the square lattice---where a fundamentally different mechanism applies---and can be as low as $U_\mathrm{c}/t\sim 10$.

The complex interplay between the kinetic frustration and the long-range interactions leads to diverse possibilities for emergent phenomena that do not exist in unfrustrated systems. This is partially reflected, for instance, in the rich phase diagram of the extended Hubbard model, with only NN interactions, on the triangular lattice. In this model,  for an electronic density $n = 2/3$ ($\delta = -1/3$) and strong NN repulsion ($V \gtrsim U/3$), the ground state is reported to be an insulating ``$200$'' ordered phase with exactly two electrons on one sublattice, while for weaker interaction strengths, it is found that two sublattices are half-filled in a hexagonal ``$110$'' order  \cite{watanabe2005charge}. For incommensurate fillings between $n = 1/2$ to $2/3$ ($\delta = -1/2$ to $-1/3$), this long-range honeycomb-type 110 charge order  (with potentially  antiferromagnetic spin correlations as $n \rightarrow 2/3$) likely coexists with metallic conductivity  \cite{tocchio2014phase}.

\subsection{Ground states}

The triangular-lattice Hubbard model with $1/r$-decaying power-law interactions is believed to describe a bevy of materials such as the charge-transfer salts $\theta$-(BEDT-TTF)$_2$X \cite{hotta2003classification} and certain layered triangular compounds, e.g., Na$_x$CoO$_2$ \cite{takada2003superconductivity}. As previously,  we start with a $4$\,$\times$\,$4$ cluster of the long-ranged Hubbard model with periodic boundaries, which we can study using exact diagonalization. The single-particle density of states on the triangular lattice exhibits a van Hove singularity  at a density of $n=1.5$ electrons per site \cite{hanisch1997lattice}. A large density of states at the Fermi level is optimal for ferromagnetism \cite{pastor1994electron,pastor1996magnetism}, so here, and throughout the rest of this section, we work with a doping of $\delta$ equaling or close to 0.5.

\begin{figure}[b]
    \centering
    \includegraphics[width=\linewidth]{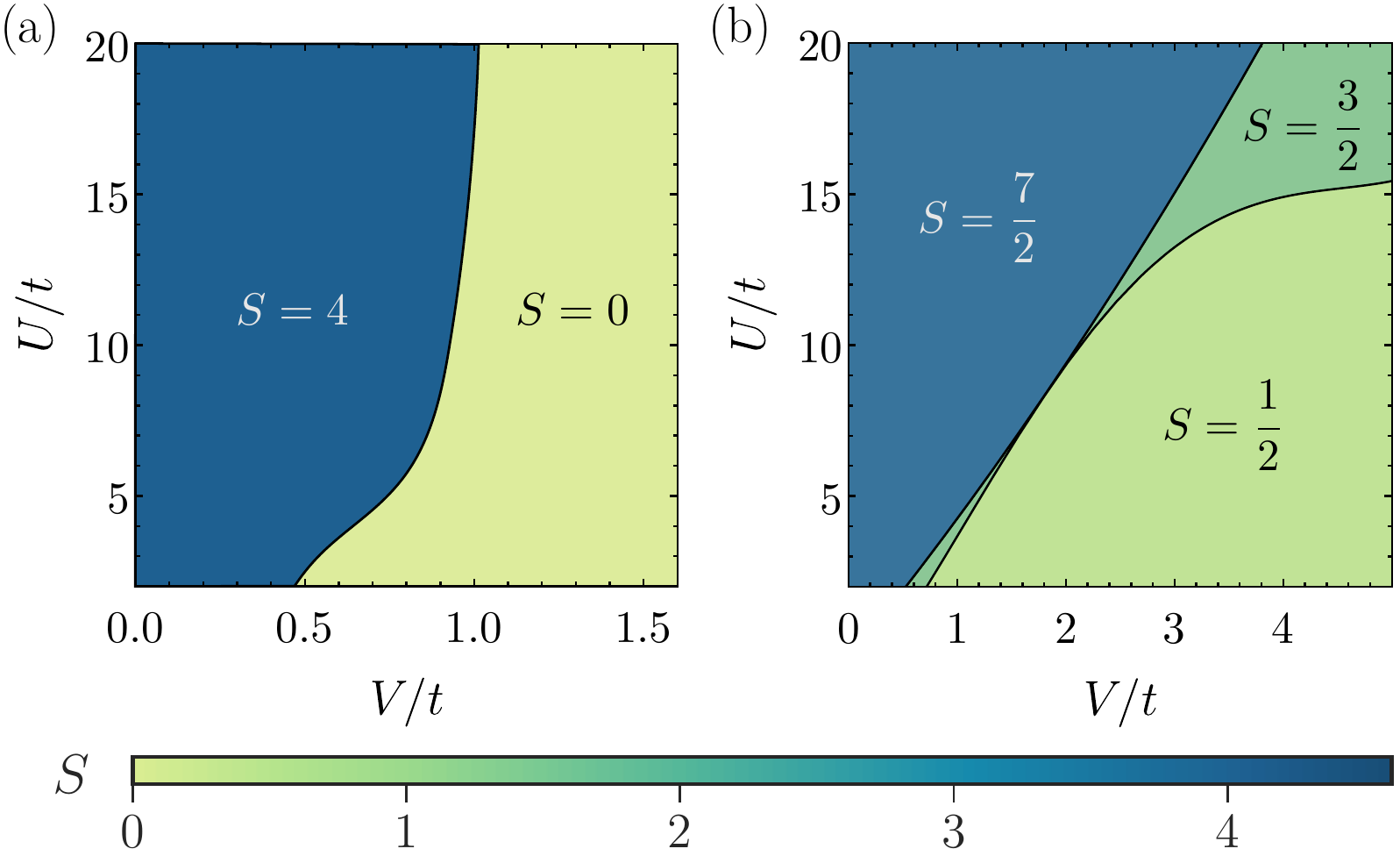}
    \caption{Total spin $S$ of $4\times 4$ triangular Hubbard plaquettes with periodic boundaries as a function of the interactions  $U/t$ and $V/t$ for a doping of (a) $\delta= 1/2$, and (b) $\delta=9/16$. Data obtained by exact diagonalization (ED) using the Lanczos algorithm.}
    \label{fig:tri_plaq}
\end{figure}

The total spin of such a cluster as a function of $U/t$, $V/t$ is shown in Fig.~\ref{fig:tri_plaq}.
For the standard Hubbard model ($V$\,$=$\,$0$), the system exhibits a ferromagnetic ground state at the dopings shown ($\delta \simeq 0.5$) even for onsite interaction strengths as little as $U/t = 2$. Introducing a finite repulsive $V$ eventually destroys this ferromagnetic phase. The details of this transition, however, are highly sensitive to the doping level. 
As shown in the two panels of Fig.~\ref{fig:tri_plaq}, which correspond to dopings of $\delta=1/2$ and $\delta=9/16$ (differing by merely $1/16$), the critical $V/t$ at which the ferromagnet--antiferromagnet phase transition occurs for a given $U/t$  can vary significantly. Additionally, for the latter doping, a  state with nonzero but low spin emerges at finite $V/t$, a feature absent in the ordinary Hubbard model.

We find that, on the triangular lattice and for the dopings considered here, the critical interaction strength $V/t$ at which ferromagnetism is destabilized exhibits only weak dependence on $U/t$. This can be understood from the fact that, in this regime, the relevant competition is primarily between the kinetic energy scale set by $t$ and the Coulomb interaction scale $V$, while the superexchange scale $J\sim t^2/U$ is comparatively small. As a result, variations in $U/t$ have a limited effect on the phase boundary. We caution, however, that this near independence is specific to the parameter range studied: at lower dopings or much larger $U/t$, where $J$ becomes more relevant, a stronger dependence on $U$ is expected.

\begin{figure}
    \centering
    \includegraphics[width=0.9\linewidth]{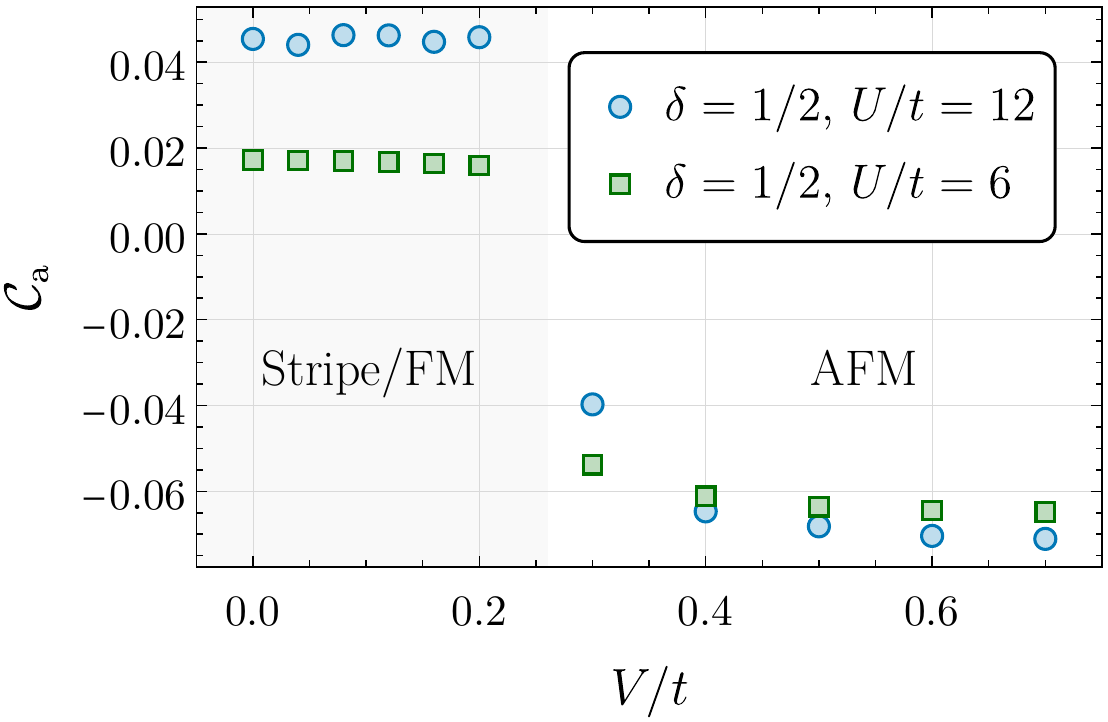}
    \caption{Ground-state nearest-neighbor spin correlations on the triangular lattice  for finite $V/t$. Together with previous work for $V$\,$=$\,$0$ \cite{samajdar2023nagaoka}, this chalks out the  phase diagram of the long-ranged Hubbard model on a $12\times 4$ cylinder. Data obtained by DMRG with maximum bond dimension $\chi=2400$.}
    \label{fig:tri_cylinder_U}
\end{figure}

As in the case of  the square lattice, we also scale the system to larger cylindrical geometries amenable to DMRG calculations. In Fig.~\ref{fig:tri_cylinder_U}, we examine the effect of a nonzero $V/t $ on these quasi-two-dimensional cylinders. The resultant behavior is noticeably simpler than for the square lattice: no partially polarized phases are observed within the investigated parameter space for the selected dopings and values of $U/t $. At $V/t \sim 0.24$, we observe a transition from the fully polarized ferromagnetic phase to the zero-spin antiferromagnet. This transition can be understood, as before, by regarding the original doped ferromagnet as a strongly interacting fluid of polarons \cite{sb2}, which is then fragmented by the effect of the Coulomb interactions that cause the individual polarons to shrink.


\subsection{Finite temperatures}

So far, we have explored the impact of long-range interactions on the ground-state magnetic properties of Hubbard-like systems. However, to apply these findings to experiments, say, with ultracold atoms or quantum dot arrays, it is crucial to understand the effect of thermal fluctuations \cite{onishi2022finite}. 

In fact, the conventional Nagaoka ferromagnet obtained by doping one hole on a square lattice is known to be rather fragile. In the limit of $U/t \rightarrow \infty$ ($J  \rightarrow 0$), the energy scale that stabilizes the fully polarized state is set by $J\propto t^2/U$ and vanishes, while the density of low-lying excitations is large, so reaching the saturated Nagaoka state requires extremely low temperatures \cite{brunner1998quantum}. This requirement precludes its observation in many experimental platforms. For example, optical-lattice Fermi-Hubbard systems can typically be cooled only to effective temperatures that are comparable to the kinetic energy scale, $t $. Given that $t \gg J$ for the square-lattice ferromagnet, we do not expect  the fully polarized phase to survive at temperatures $T$\,$\sim$\,$t $. However, on the triangular lattice, kinetic ferromagnetism is a high-temperature phenomenon since it is governed by $t$-scale (rather than $J$-scale) physics, as described above.

Motivated by these considerations, we now ask about the effects of the Coulomb repulsion at nonzero temperatures in the long-ranged triangular-lattice Hubbard model. Here, we probe such finite-temperature properties using a tensor-network algorithm based on minimally entangled typical thermal states (METTS).  For further details and benchmarking of this  method, we direct the reader to Ref.~\onlinecite{wietek2021stripes} and Appendix \ref{app:metts}.


\begin{figure}[b]
    \centering
    \includegraphics[width=\linewidth]{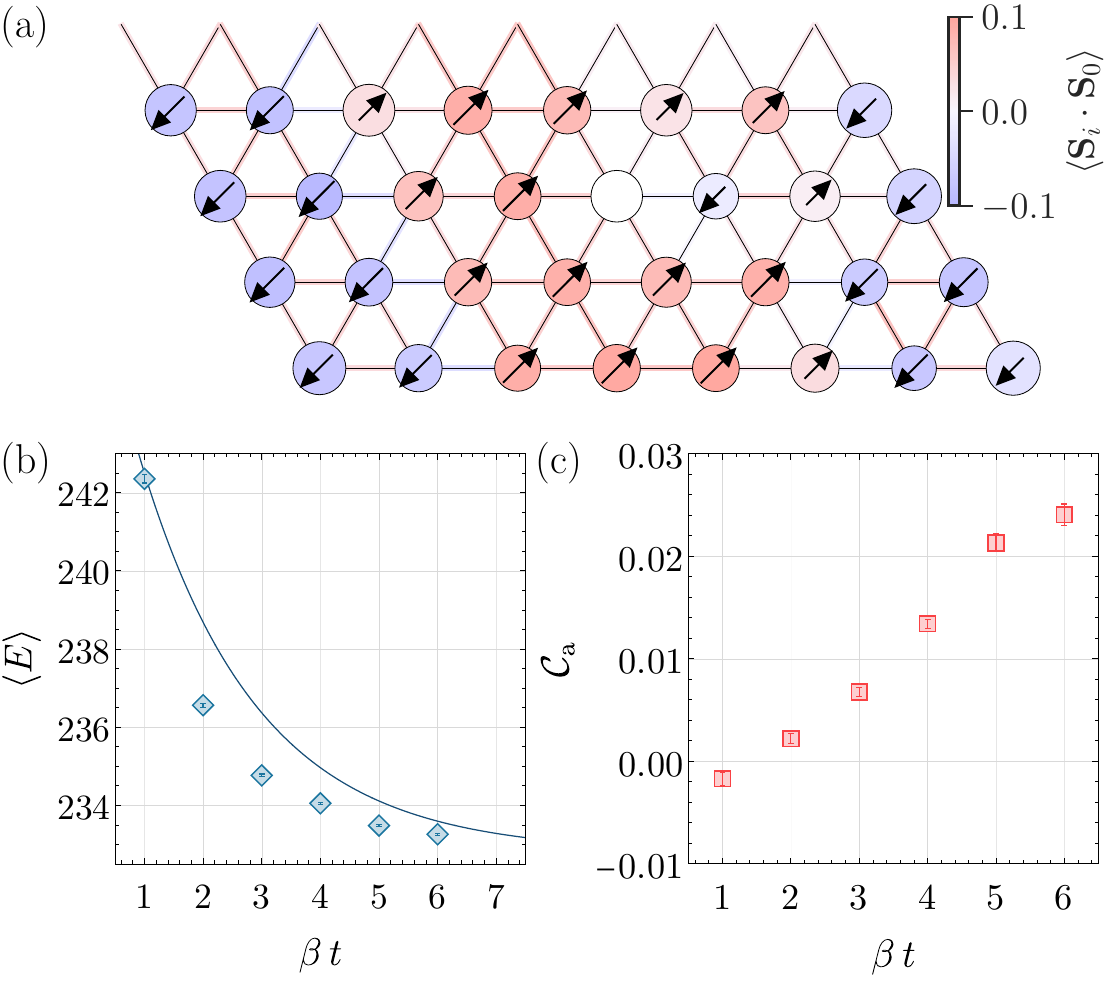}
    \caption{(a) Spin-spin correlations of a typical METTS state for the long-ranged triangular-lattice Hubbard model, at a temperature $T/t=0.2 $. Here, the parameters $U/t=12$, $V/t=0.1$ are chosen such that the ground state is a saturated ferromagnet. The colors have the same meaning as in Fig.~\ref{fig:sample_square}.     (b) Average internal energy and (c) nearest-neighbor spin correlator $\mathcal{C}_\mathrm{a}$ as a function of inverse temperature $\beta= 1/T$. The blue curve in (b) plots the analytical estimate of the energy based on single spin flips, as obtained from Eq.~\eqref{eq:spinflip}. Data in (b) and (c) are averaged over $100$ METTS samples; errorbars indicate the corresponding statistical uncertainty.}
    \label{fig:metts_tria_V0.05}
\end{figure}


Figure~\ref{fig:metts_tria_V0.05} presents our results for such a METTS calculation at finite temperatures for $V/t = 0.1$, where, per Fig.~\ref{fig:tri_cylinder_U}, the ground state is still a saturated ferromagnet. The finite-temperature characteristics of this state can be observed in Fig.~\ref{fig:metts_tria_V0.05}(a), which plots the magnetic correlations of a single representative METTS realization at $T$\,$=$\,$0.2t $: we see that ferromagnetic correlations are apparent, despite the fully polarized Nagaoka state being destroyed by thermal fluctuations.

We also compute the expectation value of the energy $\langle E \rangle$ as a function of the inverse temperature $\beta = 1 / T$, as shown in Fig.~\ref{fig:metts_tria_V0.05}(b). This numerically determined internal energy can be compared against a simple estimate as follows. 
Let  the ground-state energy of the fully polarized ferromagnetic state, as yielded by DMRG, be $E_0$. Then, taking into account single uncorrelated spin-flip excitations on top of this ferromagnetic background, we find that
\begin{align}
    \langle E\rangle^{}_\beta&=\mathrm{Tr}\left({H e^{-\beta H}}\right)\nonumber\\
    &\approx \frac{1}{Z} E^{}_0 e^{-\beta E_0}+\sum_{x,y}^{ }\left (E^{}_0+z \frac{t^2}{U}\right)e^{-\beta E^{}_0 -\beta z\frac{t^2}{U}}\nonumber\\
    &\approx E_0 \left(1+L\cdot W \frac{zt^2}{U E_0} e^{-\beta \frac{zt^2}{U}}\right),
    \label{eq:spinflip}
\end{align}
where the energy per spin excitation is assumed to be $t^2/U$, and the coordination number $z$ counts the number of neighbors of each lattice site ($z=6$ for the triangular lattice).
This approximate calculation holds only for low enough temperatures $T$\,$<$\,$z\, t^2/U$, such that only single-spin excitations are relevant and we can disregard multispin-flip excitations. Nonetheless, in Fig.~\ref{fig:metts_tria_V0.05}(b), we see that the rough analytical estimate of Eq.~\eqref{eq:spinflip} compares favorably to the numerical datapoints obtained from the METTS method over a broad range of $\beta$.

Lastly, we examine the previously introduced nearest-neighbor correlator $\mathcal{C}_\mathrm{a}$ at finite temperatures in Fig.~\ref{fig:metts_tria_V0.05}(c).
Evidently, the NN spin correlations remain positive for a nontrivial range of temperatures up to $T / t \sim 0.3 $. This provides a natural explanation for the observation of individual ferromagnetic polarons in cold-atom experiments \cite{xu2023frustration,lebrat2023observation,prichard2023directly}, in which the effective temperatures are significantly higher than the spin coupling $J \sim t^2 / U $.

\section{Discussion and outlook}

In this work, we examine, using a combination of theoretical analysis and detailed numerical simulations, the ground states and finite-temperature behavior of a  Hubbard model in which itinerant electrons interact via a $1/r$-decaying long-range Coulomb potential $V$ in addition to a strong onsite repulsion $U$. This problem had been previously considered for different plaquette geometries constructed by arranging a few ($\sim 10$) quantum dots in a controlled manner \cite{buterakos2019ferromagnetism,buterakos2023certain,buterakos2023magnetic}. In contrast, our numerics here, based on large-scale exact diagonalization and DMRG, uncover the magnetic properties of an extended many-body system, with a macroscopic number of electrons and lattice sites. In particular, we find that the fully spin-polarized Nagaoka ferromagnet remains stable up to a finite strength of $V$, beyond which it is destroyed by the onset of  phase separation, and subsequently, stripe ordering induced by the long-range interactions. 

The truncation of the Coulomb interaction to a finite range raises the question of how our results would be modified in the presence of a fully long-range $1/r$ potential. Qualitatively, including longer-distance interactions would further penalize macroscopic charge segregation, thereby enhancing the tendency toward more homogeneous or modulated charge distributions, such as domain-wall arrays or stripe phases. However, the fundamental mechanism identified here---the competition between kinetic energy, which favors delocalization and ferromagnetism, and Coulomb interactions, which penalize charge accumulation---remains unchanged. We therefore expect the qualitative structure of the phase diagram to be robust, although quantitative phase boundaries and characteristic length scales may shift. Treating truly long-range interactions within DMRG poses additional challenges due to increased entanglement and MPO complexity; recent advances in MPO compression techniques \cite{parker2020local} offer promising routes to address these effects in future work.

Irrespective of the precise phase boundaries between these competing magnetic orders in the true thermodynamic limit (which is inaccessible to DMRG), our calculations point to the leading instabilities of the half-metallic ferromagnetic state. This tendency towards local phase separation into dopant-rich ferromagnetic and dopant-poor antiferromagnetic regions \cite{visscher1974phase,emery1990phase} has been extensively investigated for the regular Hubbard and $t$-$J$ models in two dimensions \cite{hellberg1995two,shih1998phase,hellberg2000green,eisenberg2002breakdown,otsuki2014superconductivity}. Today, the numerical evidence suggests that even in the infinite-$U$ (i.e., $J$\,$\rightarrow$\,$0$) limit, the hole-doped Hubbard model phase separates for an electronic density $3/4 \le n \le 4/5$  on the square lattice \cite{liu2012phases}. 
Typically, the mechanism for phase separation is an effective short-range attraction between two doublons, driven by the fact that the merger of two adjacent polarons allows for the delocalization of the doublon core of each over a ferromagnetic region that is twice as large. Consequently, the ferromagnetic polarons tend to agglomerate (only weakly countered by the Fermi pressure of the polaron gas), leading to phase separation in the system \cite{arovas2022hubbard}. However, in our long-ranged Hubbard model, such a process is hindered by the Coulomb interaction, which acts repulsively between the excess electronic density associated with each polaron. Instead, the phase separation should  be interpreted as a consequence of the classical redistribution of charge, as is also observed in the 1D Hubbard model with NN intersite interaction \cite{sudbo1992charge}. Importantly, the competition between a local tendency to phase separation and the long-range Coulomb repulsion between doublons results in varied phenomenology that \citet{emery1993frustrated} term ``frustrated
phase separation''; this can yield an interesting variety of intermediate-scale structures, including arrays of domain walls \cite{carlson1998doped}, as observed in our striped configurations.

The physics of the long-ranged Hubbard model described in our work can be studied in arrays of gate-defined single- or few-dopant semiconductor quantum dots. Quantum simulators based on such quantum dots can reach significantly lower temperatures than achievable with ultracold atoms today, with recent experiments already having demonstrated the feasibility of cooling down to temperatures $\sim  0.02\, t$ \cite{hensgens2017quantum}. Moreover, modern fabrication methods can engineer arrays with thousands of  Coulomb-confined quantum dots arranged to subnanometer precision. This enables the controllable study of large Hubbard models---far beyond the scope of any classical simulation methods---over a broad range of model parameters tuned by the interdot separation as well as the geometry and area of the quantum dots. This flexibility could also be used to realize high-spin ground states by engineering explicitly electron-hole-asymmetric models featuring a weakly bound doublon state \cite{samajdar2023nagaoka,sb2,nielsen2007nanoscale,nielsen2010search}. On the other hand, the readout of spin degrees of freedom is more challenging for such solid-state setups compared to quantum gas microscopes, which provide direct ``snapshots'' of spin configurations \cite{mazurenko2017cold}. Potentially fruitful ideas to 
distinguish between ferromagnetic and antiferromagnetic spin textures in this regard could include measuring the nonequilibrium current shot noise \cite{ludwig2020current,brataas2020current}, magnetotransport properties \cite{telford2022coupling}, or the anomalous Hall response stemming from the intrinsic magnetization of a high-spin state \cite{abanin2009nonlocal}, as well as probing the underlying magnon excitations via spin-transfer torques \cite{chumak2015magnon,brataas2012current,roy2023spin}. Together, these possibilities present a rich variety of promising directions for future investigation.

Finally, we emphasize that the mechanisms underlying ferromagnetism can differ significantly between lattice geometries. On the square lattice, ferromagnetic tendencies at strong coupling are closely related to Nagaoka-type physics, which relies on the bipartite structure of the lattice. In contrast, the triangular lattice is nonbipartite, and any ferromagnetic tendencies arise from different kinetic and band-structure effects. Our inclusion of both lattices is intended to highlight how lattice geometry influences the interplay between kinetic energy and long-range Coulomb interactions. In particular, the triangular lattice results should not be interpreted as direct realizations of Nagaoka ferromagnetism, but rather as an exploration of how similar energetic competitions manifest in a frustrated, nonbipartite setting. Investigating related scenarios such as the physics of van Hove singularities or models with additional hopping terms (e.g., square lattices with a finite next-nearest-neighbor hopping $t'$) represents an interesting direction for future work.

\begin{acknowledgments}
We thank M.B. Donnelly, S.K. Gorman, and M.Y. Simmons for useful discussions. J.D. acknowledges support from the German Research Foundation (DFG) through the Collaborative Research Center, Project ID 314695032 SFB 1277 (project A03) and
the German Academic Scholarship Foundation. R.S. and R.N.B. acknowledge the hospitality of the Aspen Center for Physics, where some of this work was performed. The participation of R.N.B. at the Aspen Center for Physics was supported in part by the National Science Foundation grant PHY-2210452, and that of R.S. by a grant from the Simons Foundation (1161654, Troyer). R.S. is supported by the Princeton Quantum Initiative Fellowship. The DMRG calculations presented in this paper were performed using the \textsc{ITensor} library \cite{ITensor} on computational resources managed and supported by Princeton Research Computing, a consortium of groups including the Princeton Institute for Computational Science and Engineering (PICSciE) and the Office of Information Technology's High Performance Computing Center and Visualization Laboratory at Princeton University. 
\end{acknowledgments}

\appendix
\section{Benchmarks of the METTS algorithm}
\label{app:metts}


In this Appendix, we provide a brief comparison of results obtained using the METTS algorithm and those computed via exact diagonalization of small samples. Since we are interested in finite-temperature properties, finding the ground states---using, for instance, the Lanczos  algorithm---does not alone suffice. While ground-state methods such as Lanczos diagonalization are well suited for zero-temperature studies, finite-temperature properties require access to a large portion of the spectrum. Finite-temperature Lanczos methods (FTLM) \cite{prelovsek2013ground,prelovsek2020,schnack2020finite,sugiura2020thermal} provide a powerful approach for small systems, where many eigenstates can be efficiently sampled. However, for the quasi-two-dimensional geometries considered here, the Hilbert space grows exponentially with system size, rendering such approaches impractical. Instead, we employ the METTS algorithm, which is specifically designed to access finite-temperature properties within a matrix-product-state framework and allows us to study thermal behavior on the same cylinder geometries used in our ground-state DMRG calculations.

\begin{figure}[H]
    \centering
    \includegraphics[width=\linewidth]{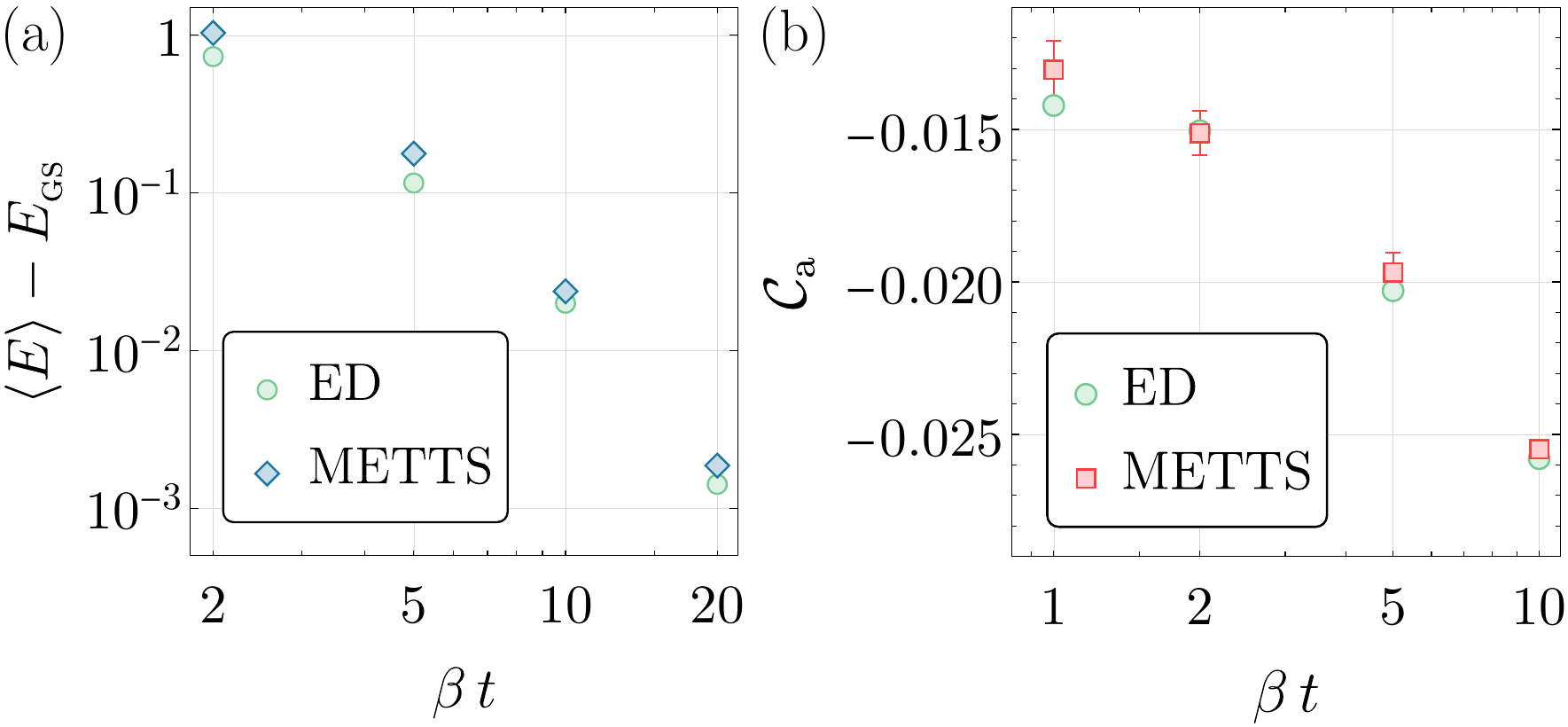}
    \caption{Comparison of the internal energy and nearest-neighbor correlations obtained using the METTS algorithm at finite temperatures with the corresponding exact diagonalization (ED) results for a small sample. The calculation is set up on a $3\times 3$ square lattice with cylindrical boundary conditions for $U/t=12$, $V=0$, and doping of $\delta=1/9$. METTS data are averaged over $100$ samples; errorbars indicate the corresponding statistical uncertainty.}
    \label{fig:metts_ed}
\end{figure}

To benchmark this methodology, we study a $3$\,$\times$\,$3$ square-lattice Hubbard model, without long-range Coulomb interactions, on a cylindrical geometry. 
First, we evaluate the internal energy for varying inverse temperature $\beta=1/T$ in Fig.~\ref{fig:metts_ed}(a): the data  obtained from averaging over 100 METTS steps tracks the exact results within statistical errorbars. Similarly, the average nearest-neighbor spin correlator, shown in Fig.~\ref{fig:metts_ed}(b), is also in reasonably good agreement with the exact diagonalization data up to errorbars. Thus, we conclude that over the range of $\beta$ considered, the finite-temperature properties of the Hubbard model can be reasonably estimated using METTS. 
Our use of the METTS algorithm closely follows Ref.~\onlinecite{wietek2021stripes}, which  also provides more detailed benchmarking.

\bibliographystyle{apsrev4-2}
\nocite{apsrev41Control}
\bibliography{refs}

\end{document}